\documentclass[12pt]{article}

\usepackage{cite}

\title{Lagrangian formulation for the infinite spin\\
$N=1$ supermultiplets in $d=4$}

\author{I.L. Buchbinder${}^{ab}$\thanks{joseph@tspu.edu.ru},
M.V. Khabarov${}^{cd}$\thanks{maksim.khabarov@ihep.ru}, T.V.
Snegirev${}^{ae}$\thanks{snegirev@tspu.edu.ru}, Yu.M.
Zinoviev${}^{cd}$\thanks{Yurii.Zinoviev@ihep.ru}
\\[0.5cm]
\it{\small ${}^a$Department of Theoretical Physics, Tomsk State
Pedagogical University,}\\
\it{\small Tomsk, 634061, Russia}\\
\it{\small ${}^b$National Research Tomsk State University, Tomsk
634050, Russia}\\
\it{\small ${}^c$Institute for High Energy Physics of National
Research Center "Kurchatov Institute"} \\
\it{\small Protvino, Moscow Region, 142281, Russia} \\
\it\small{ ${}^d$Moscow Institute of Physics and Technology (State
University),} \\
\it{\small Dolgoprudny, Moscow Region, 141701, Russia}\\
\it{\small ${}^e$National Research Tomsk Polytechnic University,
Tomsk 634050, Russia}}

\date{}

\begin{document}

\maketitle

\begin{abstract}
We provide an explicit  Lagrangian construction for the massless
infinite spin $N=1$ supermultiplet in four dimensional Minkowski
space. Such a supermultiplet contains a pair of massless bosonic and
a pair of massless fermionic infinite spin fields with properly
adjusted dimensionful parameters. We begin with the gauge invariant
Lagrangians for such massless infinite spin bosonic and fermionic
fields and derive the supertransformations which leave the sum of
their Lagrangians invariant. It is shown that the algebra of these
supertransformations is closed on-shell.

\end{abstract}

\thispagestyle{empty}
\newpage
\setcounter{page}{1}

\section{Introduction}

Infinite spin fields (see \cite{BKRX02,BS17} and references therein)
have attracted a lot of interest recently. A number of different
approaches for their description were proposed
\cite{BM05,Ben13,ST14,ST14a,Riv14,BNS15,Met16,Met17,Zin17,MN17,KhZ17,
Met17b,AG17,Met18,BFIR18,BKT18,Riv18,ACG18,BFI19,BGK19}. Also the
investigations of possible interactions for such fields started
\cite{met17a,BMN17,Met18b}. At the same time there exists just a few
results on the supermultiplets containing such particles
\cite{Zin17,BFI19,BGK19}, while their classification is already known
for a long time \cite{BKRX02}.

One of the interesting features of the infinite spin fields is that
being massless they however depend on some dimensionful parameter,
related with the value of the second Casimir operator of Poincare
group. In many respects such fields can be considered as a limit of
massive higher spin fields, where mass $m \to 0$ and spin $s \to
\infty$ so that $\mu = ms \to const$. In-particular, it appears that
the gauge invariant formalism for the description of massive higher
spin fields \cite{Zin01,Met06,Zin08b,PV10} can be successfully applied
for the description of infinite spin fields as well
\cite{Met16,Met17,Zin17,KhZ17}. Taking into account that such gauge
invariant formalism remains to be the only effective way to construct
massive higher spin supermultiplets
\cite{Zin07a,BSZ17a,BKhSZ19,BKhSZ19a}, it seems natural to apply this
approach to the construction of the infinite spin supermultiplets.

In this note we provide an explicit on-shell Lagrangian realization
for the massless $N=1$ infinite spin supermultiplet in flat four
dimensional Minkowski space. Our paper is organzed as follows. In
Section 2 we give the gauge invariant Lagrangians for the massless
bosonic and fermionic infinite spin fields. Then in Section 3 we
find supertransformations which leave the sum of their free
Lagrangians invariant and such that the algebra of the
supertransformations is closed on-shell.

\noindent
{\bf Notations and conventions} We use the same frame-like multispinor
formalism as in \cite{BKhSZ19}, where all objects are one or zero
forms having some number of dotted $\dot\alpha=1,2$ and undotted
$\alpha=1,2$ completely symmetric indices. Coordinate-free description
of flat Minkowski space is given by the external derivative $d$ and a
background one-form frame $e^{\alpha\dot\alpha}$ as well as by basic
two, three and four forms:
$$
E_2{}^{\alpha\beta},\quad E_2{}^{\dot\alpha\dot\beta},\quad
E_3{}^{\alpha\dot\alpha},\quad E_4,
$$
which are defined as follows:
\begin{eqnarray*}
e^{\alpha\dot\alpha}\wedge  e^{\beta\dot\beta} &=&
\varepsilon^{\alpha\beta} E^{\dot\alpha\dot\beta} +
\varepsilon^{\dot\alpha\dot\beta }E^{\alpha\beta},
\\
E^{\alpha\beta}\wedge  e^{\gamma\dot\alpha} &=&
\varepsilon^{\alpha\gamma} E^{\beta\dot\alpha}+
\varepsilon^{\beta\gamma} E^{\alpha\dot\alpha},
\\
E^{\dot\alpha\dot\beta}\wedge  e^{\alpha\dot\gamma} &=&
-\varepsilon^{\dot\alpha\dot\gamma} E^{\alpha\dot\beta}
-\varepsilon^{\dot\beta\dot\gamma} E^{\alpha\dot\alpha},
\\
E^{\alpha\dot\alpha}\wedge  e^{\beta\dot\beta} &=&
\varepsilon^{\alpha\beta} \varepsilon^{\dot\alpha\dot\beta}E.
\end{eqnarray*}

\section{Kinematics of infinite spin fields}

In this work we use a description for the infinite spin bosonic and
fermionic fields based on the gauge invariant formalism for the
massive higher spin fields \cite{Zin01,Met06,Zin08b,PV10}, which has
already been successfully applied for the infinite spin fields as
well \cite{Met16,Met17,Zin17,KhZ17}.

\subsection{Infinite spin boson}

An infinite spin bosonic field  in $d=4$ contains an infinite number
of helicities $0 \le l < \infty$, so for the gauge invariant
formulation we introduce a number of physical and auxiliary
one-forms $f^{\alpha(k)\dot{\alpha}(k)}$,
$\Omega^{\alpha(k+1)\dot{\alpha}(k-1)} + h.c.$, $1 \le k < \infty$,
as well as zero-forms $B^{\alpha(2)} + h.c.$ and one-form $A$ for
the helicities $\pm1$, while helicity 0 is described by two
zero-forms $\pi^{\alpha\dot{\alpha}}$ and $\varphi$. The general
ansatz for the corresponding bosonic Lagrangian ${\cal L}_B$ is
written as follows:
\begin{eqnarray}
\frac{1}{i}{\cal L}_B &=& \sum_{k=1}^{\infty} (-1)^{k+1} [ (k+1)
\Omega^{\alpha(k)\beta\dot{\alpha}(k-1)} E_\beta{}^\gamma
\Omega_{\alpha(k)\gamma\dot{\alpha}(k-1)}\nonumber\\
&& - (k-1) \Omega^{\alpha(k+1)\dot{\alpha}(k-2)\dot{\beta}}
E_{\dot{\beta}}{}^{\dot{\gamma}}
\Omega_{\alpha(k+1)\dot{\alpha}(k-2)\dot{\gamma}} + 2
\Omega^{\alpha(k)\beta\dot{\alpha}(k-1)} e_\beta{}^{\dot{\beta}} d
f_{\alpha(k)\dot{\alpha}(k-1)\dot{\beta}} + h.c. ] \nonumber
 \\
 && + 4 [ E B^{\alpha(2)} B_{\alpha(2)} + E
B^{\dot{\alpha}(2)} B_{\dot{\alpha}(2)} ] + 2
[ E^{\alpha(2)} B_{\alpha(2)} d A - E^{\dot{\alpha(2)}}
B_{\dot{\alpha}(2)} d A ] \nonumber
 \\
 && - 6 E \pi^{\alpha,\dot{\alpha}}
\pi_{\alpha,\dot{\alpha}} - 12 E^{\alpha,\dot{\alpha}}
\pi_{\alpha,\dot{\alpha}} d \varphi \nonumber
 \\
 && + \sum_{k=1}^{\infty} (-1)^{k+1}  a_k [
\Omega^{\alpha(k)\beta(2)\dot{\alpha}(k)} E_{\beta(2)}
f_{\alpha(k)\dot{\alpha}(k)}\nonumber\\
&& + \frac{k}{(k+2)} \Omega^{\alpha(k+1)\dot{\alpha}(k-1)}
E^{\dot{\beta}(2)} f_{\alpha(k+1)\dot{\alpha}(k-1)\dot{\beta}(2)} +
h.c.] \nonumber
 \\
 && + a_0 [ \Omega^{\alpha(2)} E_{\alpha(2)} A -
\Omega^{\dot{\alpha}(2)} E_{\dot{\alpha}(2)} A ]
 - 2a_0 [ B^{\alpha\beta} E_\alpha{}^{\dot{\alpha}}
f_{\beta,\dot{\alpha}} + B^{\dot{\alpha}\dot{\beta}}
E^\alpha{}_{\dot{\alpha}} f_{\alpha,\dot{\beta}} ]
 + \tilde{a}_0 E^{\alpha\dot{\alpha}} \pi_{\alpha,\dot{\alpha}} A
\nonumber
 \\
 && + \sum_{k=1}^{\infty} (-1)^{k+1} [ b_k
f^{\alpha(k-1)\beta\dot{\alpha}(k)} E_\beta{}^\gamma
f_{\alpha(k-1)\gamma\dot{\alpha}(k)} + h.c. ] +
\frac{a_0\tilde{a}_0}{2} E^{\alpha\dot{\alpha}}
f_{\alpha,\dot{\alpha}} \varphi + 3a_0{}^2 E \varphi^2 \label{lagb}.
\end{eqnarray}
Here $a_k,b_k$ are arbitrary dimensional coefficients providing the
mixture of the different helicities into one multiplet. The
Lagrangian (\ref{lagb}) has a common structure for the massive
higher spin gauge invariant description, namely it contains the
usual kinetic and mass-like terms for all the helicity components as
well as the cross-terms connecting the nearest neighbours. Such a
structure follows from the requirement that we still must have all
(appropriately modified) gauge symmetries, which our helicity
components initially possessed. The ansatz for these modified gauge
transformations (consistent with the structure of the ansatz for the
Lagrangian ${\cal L}_B$) has the form:
\begin{eqnarray}
\delta \Omega^{\alpha(k+1)\dot{\alpha}(k-1)} &=& d
\eta^{\alpha(k+1)\dot{\alpha}(k-1)} + e_\beta{}^{\dot{\alpha}}
\eta^{\alpha(k+1)\beta\dot{\alpha}(k-2)}
 + \frac{a_k}{2} e_{\beta\dot{\beta}}
\eta^{\alpha(k+1)\beta\dot{\alpha}(k-1)\dot{\beta}} \nonumber
 \\
 && + \frac{a_{k-1}}{(k+1)(k+2)} e^{\alpha\dot{\alpha}}
\eta^{\alpha(k)\dot{\alpha}(k-2)}
 + \frac{b_k}{2(k+1)} e^\alpha{}_{\dot{\beta}}
\xi^{\alpha(k)\dot{\alpha}(k-1)\dot{\beta}}, \nonumber
 \\
\delta f^{\alpha(k)\dot{\alpha}(k)} &=& d
\xi^{\alpha(k)\dot{\alpha}(k)} + e_\beta{}^{\dot{\alpha}}
\eta^{\alpha(k)\beta\dot{\alpha}(k-1)} + e^\alpha{}_{\dot{\beta}}
\eta^{\alpha(k-1)\dot{\alpha}(k)\dot{\beta}} \nonumber
 \\
 && + \frac{ka_k}{2(k+2)} e_{\beta\dot{\beta}}
\xi^{\alpha(k)\beta\dot{\alpha}(k)\dot{\beta}} +
\frac{a_{k-1}}{2k(k+1)}  e^{\alpha\dot{\alpha}}
\xi^{\alpha(k-1)\dot{\alpha}(k-1)}, \nonumber
 \\
\delta \Omega^{\alpha(2)} &=& d \eta^{\alpha(2)} + \frac{a_1}{2}
e_{\beta\dot{\beta}} \eta^{\alpha(2)\beta\dot{\beta}} + \frac{b_1}{4}
e^\alpha{}_{\dot{\beta}} \xi^{\alpha\dot{\beta}},
 \\
\delta f^{\alpha\dot{\alpha}} &=& d \xi^{\alpha\dot{\alpha}} +
e_\beta{}^{\dot{\alpha}} \eta^{\alpha\beta} +
e^\alpha{}_{\dot{\beta}} \eta^{\dot{\alpha}\dot{\beta}}
+ \frac{a_1}{6} e_{\beta\dot{\beta}}
\xi^{\alpha\beta\dot{\alpha}\dot{\beta}} - \frac{a_0}{4}
e^{\alpha\dot{\alpha}} \xi, \nonumber
 \\
\delta B^{\alpha(2)} &=& \frac{a_0}{2} \eta^{\alpha(2)}, \qquad
\delta A = d \xi - \frac{a_0}{2} e^{\alpha\dot{\alpha}}
\xi_{\alpha\dot{\alpha}}, \nonumber
 \\
\delta \pi^{\alpha\dot{\alpha}} &=& - \frac{a_0\tilde{a}_0}{24}
\xi^{\alpha\dot{\alpha}}, \qquad \delta \varphi =
\frac{\tilde{a}_0}{12} \xi. \nonumber
\end{eqnarray}

The invariance of the Lagrangian under these gauge transformations
leads to the following relations on the parameters:
$$
(k+2)b_k = (k-1)b_{k-1},
$$
$$
\frac{k}{4(k+2)}a_k{}^2 - \frac{(k-1)}{4(k+1)}a_{k-1}{}^2 +
b_k = 0,
$$
$$
\frac{1}{12}a_1{}^2 - \frac{1}{4}a_0{}^2 + b_1 - 4\lambda^2 = 0,
\qquad \tilde{a}_0{}^2 = 72b_1.
$$
A general solution for these relations depends on two parameters. We
choose $a_0$ and $b_1$ and obtain:
\begin{equation}
a_k{}^2 = \frac{(k+2)}{k} [ a_0{}^2 - \frac{6k(k+3)}{(k+1)(k+2)}b_1],
\qquad b_k = \frac{6b_1}{k(k+1)(k+2)}.
\end{equation}
For the resulting Lagrangian be hermitian (and the theory to be
unitary) we must have $a_k{}^2 \ge 0$ for all $k$. This leads to the
two types of solutions \cite{Met06,KhZ17}:
\begin{itemize}
\item The solutions with the whole spectrum of helicities
$0 \le k < \infty$, which requires $a_0{}^2 \ge 6b_1$. For
$a_0{}^2 > 6b_1$ they correspond to the tachyonic fields, while for
$a_0{}^2 = 6b_1$ we obtain a massless infinite spin boson (the case we
are mostly interested in):
\begin{equation}
a_k{}^2 = \frac{2a_0{}^2}{k(k+1)}, \qquad
b_k = \frac{a_0{}^2}{k(k+1)(k+2)}. \label{parb}
\end{equation}
\item The second type of solutions have the spectrum
$s \le l < \infty$
$$
a_s = 0 \quad \Rightarrow \quad
a_k{}^2 = - \frac{12(k+2)(k+s+3)(k-s)}{k(k+1)(s+1)(s+2)} b_1.
$$
The positivity of $a_k{}^2$ requires $b_1$ to be negative, so all
these solutions are tachyonic.
\end{itemize}

To simplify our construction of the supermultiplet we do not introduce
any supertransformations for the auxiliary fields $\Omega$, $b$ and
$\pi$. Instead, all calculations are done up to the terms proportional
to the auxiliary field equations of motion that is equivalent to the
following "zero torsion conditions":
\begin{eqnarray}
{\cal T}^{\alpha(k)\dot\alpha(k)} &=& d f^{\alpha(k)\dot\alpha(k)} +
e_\beta{}^{\dot\alpha} \Omega^{\alpha(k)\beta\dot\alpha(k-1)} +
e^\alpha{}_{\dot\beta} \Omega^{\alpha(k-1)\dot\alpha(k)\dot\beta}
\nonumber
 \\
 && + \frac{ka_k}{2(k+2)} e_{\beta\dot\beta}
f^{\alpha(k)\beta\dot\alpha(k)\dot\beta} + \frac{a_{k-1}}{2k(k+1)}
e^{\alpha\dot\alpha} f^{\alpha(k-1)\dot\alpha(k-1)} \approx 0,
\nonumber
 \\
{\cal T}^{\alpha\dot\alpha} &=& d f^{\alpha\dot\alpha} +
e_\beta{}^{\dot\alpha} \Omega^{\alpha\beta} + e^\alpha{}_{\dot\beta}
\Omega^{\dot\alpha\dot\beta} + \frac{a_1}{6} e_{\beta\dot\beta}
f^{\alpha\beta\dot\alpha\dot\beta} - \frac{a_0}{4}
e^{\alpha\dot\alpha} A \approx 0,
 \\
{\cal T} &=& d A + 2(E_{\alpha(2)} B^{\alpha(2)} + E_{\dot\alpha(2)}
B^{\dot\alpha(2)} - \frac{a_0}{2} e_{\alpha\dot\alpha}
f^{\alpha\dot\alpha} \approx 0, \nonumber \label{zerotor}
 \\
{\cal C} &=& d \varphi + e_{\alpha\dot\alpha} \pi^{\alpha\dot\alpha} -
\frac{\tilde{a}_0}{12} A \approx 0. \nonumber
\end{eqnarray}
As for the supertransformations for the physical fields $f$, $A$ and
$\varphi$, the corresponding variation of the Lagrangian can be
compactly written as follows
\begin{eqnarray}
\delta {\cal L}_B &=& 2i \sum_{k=1}^{\infty} (-1)^k {\cal
R}^{\alpha(k)\beta\dot\alpha(k-1)} e_\beta{}^{\dot\beta} \delta
f_{\alpha(k)\dot\alpha(k-1)\dot\beta} \nonumber
 \\
 && - 2i E_{\alpha(2)} {\cal C}^{\alpha(2)} \delta A + 12i
E_{\alpha\dot\alpha} {\cal C}^{\alpha\dot\alpha} \delta \varphi + h.c.
\label{varb}
\end{eqnarray}
where we introduced "curvatures":
\begin{eqnarray}
{\cal R}^{\alpha(k+1)\dot\alpha(k-1)} &=& d
\Omega^{\alpha(k+1)\dot\alpha(k-1)} + e_\beta{}^{\dot\alpha}
\Omega^{\alpha(k+1)\beta\dot\alpha(k-2)} + \frac{a_k}{2}
e_{\beta\dot\beta} \Omega^{\alpha(k+1)\beta\dot\alpha(k-1)\dot\beta}
\nonumber
 \\
 && + \frac{a_{k-1}}{(k+1)(k+2)} e^{\alpha\dot\alpha}
\Omega^{\alpha(k)\dot\alpha(k-2)} + \frac{b_k}{2(k+1)}
e^\alpha{}_{\dot\beta} f^{\alpha(k)\dot\alpha(k-1)\dot\beta},
\nonumber
 \\
{\cal R}^{\alpha(2)} &=& d \Omega^{\alpha(2)} + \frac{a_1}{2}
e_{\beta\dot\beta} \Omega^{\alpha(2)\beta\dot\beta} + \frac{b_1}{4}
e^\alpha{}_{\dot\beta} f^{\alpha\dot\beta}
 - \frac{a_0}{4} E^\alpha{}_\beta B^{\alpha\beta} +
\frac{a_0\tilde{a}_0}{24} E^{\alpha(2)} \varphi,
 \\
{\cal C}^{\alpha(2)} &=& d B^{\alpha(2)} - \frac{a_0}{2}
\Omega^{\alpha(2)} - \frac{\tilde{a}_0}{24} e^\alpha{}_{\dot\beta}
\pi^{\alpha\dot\beta}, \nonumber
 \\
{\cal C}^{\alpha\dot\alpha} &=& d \pi^{\alpha\dot\alpha} +
\frac{a_0\tilde{a}_0}{24} f^{\alpha\dot\alpha} -
\frac{\tilde{a}_0}{12} (e_\beta{}^{\dot\alpha} B^{\alpha\beta} +
e^\alpha{}_{\dot\beta} B^{\dot\alpha\dot\beta}) + \frac{a_0{}^2}{8}
e^{\alpha\dot\alpha} \varphi. \nonumber
\end{eqnarray}

\subsection{Infinite spin fermion}

For the gauge invariant description of the infinite spin fermionic
field we need one-forms $\Phi^{\alpha(k+1)\dot{\alpha}(k)}$,
$\Phi^{\alpha(k)\dot{\alpha}(k+1)}$, $0 \le k < \infty$, as well as
zero-forms $\phi^\alpha$, $\phi^{\dot{\alpha}}$. The general ansatz
for the corresponding Lagrangian ${\cal L}_F$ is written written as
follows:
\begin{eqnarray}
{\cal L}_F &=& \sum_{k=0}^{\infty} (-1)^{k+1}
\Phi_{\alpha(k)\beta\dot{\alpha}(k)} e^\beta{}_{\dot{\beta}} d
\Phi^{\alpha(k)\dot{\alpha}(k)\dot{\beta}} - \phi_\alpha
E^\alpha{}_{\dot{\alpha}} d \phi^{\dot{\alpha}} \nonumber
 \\
 && + \sum_{k=1}^{\infty} (-1)^{k+1} c_k
\Phi_{\alpha(k-1)\beta(2)\dot{\alpha}(k)} E^{\beta(2)}
\Phi^{\alpha(k-1)\dot{\alpha}(k)} + c_0 \Phi_\alpha
E^\alpha{}_{\dot{\alpha}} \phi^{\dot{\alpha}} + h.c. \nonumber
 \\
 && + \sum_{k=0}^{\infty} (-1)^{k+1} d_k [ (k+2)
\Phi_{\alpha(k)\beta\dot{\alpha}(k)} E^\beta{}_\gamma
\Phi^{\alpha(k)\gamma\dot{\alpha}(k)} - k
\Phi_{\alpha(k+1)\dot{\alpha}(k-1)\dot{\beta}}
E^{\dot{\beta}}{}_{\dot{\gamma}}
\Phi^{\alpha(k+1)\dot{\alpha}(k-1)\dot{\gamma}} + h.c. ] \nonumber
 \\
 && + 2d_0 E \phi_\alpha \phi^\alpha + h.c. \label{lagf}
\end{eqnarray}
Here $c_k,d_k$ are the dimensionful coefficients providing the
mixture of the different helicities into one multiplet. The
Lagrangian (\ref{lagf}) has the same common structure as in the
bosonic case. The ansatz for the supertransformations (consistent
with that for the Lagrangian) ${\cal L}_F$) has the form:
\begin{eqnarray}
\delta \Phi^{\alpha(k+1)\dot{\alpha}(k)} &=& d
\eta^{\alpha(k+1)\dot{\alpha}(k)} + c_{k+1} e_{\beta\dot{\beta}}
\eta^{\alpha(k+1)\beta\dot{\alpha}(k)\dot{\beta}} + 2d_k
e^\alpha{}_{\dot{\beta}} \eta^{\alpha(k)\dot{\alpha}(k)\dot{\beta}} +
\frac{c_k}{k(k+2)} e^{\alpha\dot{\alpha}}
\eta^{\alpha(k)\dot{\alpha}(k-1)}, \nonumber
 \\
\delta \Phi^{\alpha(k)\dot{\alpha}(k+1)} &=& d
\eta^{\alpha(k)\dot{\alpha}(k+1)} + c_{k+1} e_{\beta\dot{\beta}}
\eta^{\alpha(k)\beta\dot{\alpha}(k+1)\dot{\beta}} + 2d_k
e_\beta{}^{\dot{\alpha}} \eta^{\alpha(k)\beta\dot{\alpha}(k)} +
\frac{c_k}{k(k+2)} e^{\alpha\dot{\alpha}}
\eta^{\alpha(k-1)\dot{\alpha}(k)}, \nonumber
 \\
\delta \Phi^\alpha &=& d \eta^\alpha + c_1 e_{\beta\dot{\beta}}
\eta^{\alpha\beta\dot{\beta}} + 2d_0 e^\alpha{}_{\dot{\beta}}
\eta^{\dot{\beta}}, \qquad
\delta \phi^\alpha = c_0 \eta^\alpha,
 \\
\delta \Phi^{\dot{\alpha}} &=& d \eta^{\dot{\alpha}} + c_1
e_{\beta\dot{\beta}} \eta^{\beta\dot{\alpha}\dot{\beta}} + 2d_0
e_\beta{}^{\dot{\alpha}} \eta^\beta, \qquad
\delta \phi^{\dot{\alpha}} = c_0 \eta^{\dot{\alpha}}. \nonumber
\end{eqnarray}
The invariance of the Lagrangian ${\cal L}_F$ under these gauge
transformations leads to the following relations on the parameters:
$$
(k+2)d_k = kd_{k-1}, \qquad k \ge 1,
$$
$$
c_{k+1}{}^2 - c_k{}^2 + 4(2k+3)d_k{}^2 = 0,
$$
$$
2c_1{}^2 - c_0{}^2 + 24d_0{}^2 = 0.
$$
General solution again depends on the two parameters. We choose $c_0$
and $d_0$ and obtain:
\begin{equation}
c_k{}^2 = \frac{c_0{}^2}{2} - \frac{16k(k+2)}{(k+1)^2}d_0{}^2, \qquad
d_k = \frac{2d_0}{(k+1)(k+2)}.
\end{equation}
As in the bosonic case we have two types of solution
\cite{Met17,KhZ17}.
\begin{itemize}
\item Solutions  with the whole spectrum of helicities
$1/2 \le l < \infty$, which requires $c_0{}^2 \ge 32d_0{}^2$. Most of
them are tachyonic, while for $c_0{}^2 = 32d_0{}^2$ we obtain a
massless infinite spin fermion
\begin{equation}
c_k{}^2 = \frac{c_0{}^2}{2(k+1)^2}, \qquad
d_k = \pm \frac{c_0}{2\sqrt{2}(k+1)(k+2)}. \label{parf}
\end{equation}
\item Solutions with the spectrum $s+1/2 \le l < \infty$
$$
c_s = 0 \quad \Rightarrow \quad
c_k{}^2 = - \frac{16(k+s+2)(k-s)}{(k+1)^2(s+1)^2}d_0{}^2,
$$
where the positivity of $c_k{}^2$ requires $d_0{}^2$ to be negative
and hence $d_0$ to be imaginary.
\end{itemize}
Note, that in the fermionic case all tachyonic solutions require
imaginary masses so that the Lagrangian is not hermitian. Thus in
what follows we restrict ourselves with the massless infinite spin
bosons and fermions.

As in the bosonic case, the variation of the Lagrangian ${\cal L}_F$
under the arbitrary transformations for the physical fields can be
compactly written as follows:
\begin{equation}
\delta {\cal L}_F = \sum_{k=0}^\infty (-1)^k {\cal
F}_{\alpha(k)\beta\dot\alpha(k)} e^\beta{}_{\dot\beta} \delta
\Phi^{\alpha(k)\dot\alpha(k)\dot\beta} - {\cal C}_\alpha
E^\alpha{}_{\dot\alpha} \delta \phi^{\dot\alpha} + h.c. \label{varf}
\end{equation}
where we introduced gauge invariant "curvatures":
\begin{eqnarray}
{\cal F}^{\alpha(k+1)\dot\alpha(k)} &=& d
\Phi^{\alpha(k+1)\dot\alpha(k)} + c_{k+1} e_{\beta\dot\beta}
\Phi^{\alpha(k+1)\beta\dot\alpha(k)\dot\beta} \nonumber
 \\
 && + 2d_k e^\alpha{}_{\dot\beta}
\Phi^{\alpha(k)\dot\alpha(k)\dot\beta} + \frac{c_k}{k(k+2)}
e^{\alpha\dot\alpha} \Phi^{\alpha(k)\dot\alpha(k-1)}, \nonumber
 \\
{\cal F}^\alpha &=& d \Phi^\alpha + c_1 e_{\beta\dot\beta}
\Phi^{\alpha\beta\dot\beta} + 2d_0 e^\alpha{}_{\dot\beta}
\Phi^{\dot\beta} - \frac{c_0}{3} E^\alpha{}_\beta \phi^\beta,
 \\
{\cal C}^\alpha &=& d \phi^\alpha - c_0 \Phi^\alpha + 2d_0
e^\alpha{}_{\dot\beta} \phi^{\dot\beta}. \nonumber
\end{eqnarray}

\section{Infinite spin supermultiplet}

In this section we construct a supermultiplet containing infinite spin
bosonic and fermionic fields. Let us consider one massless infinite
spin boson with the Lagrangian (\ref{lagb}) and the parameters
(\ref{parb}) and one massless infinite spin fermion with the
Lagrangian (\ref{lagf}) and the parameters (\ref{parf}). Taking into
account close similarity between the gauge invariant description for
massive finite spin fields and massless infinite spin ones, we take
the same general ansatz for the supertransformations as in
\cite{BKhSZ19}. Namely, for the bosonic components we take
\begin{eqnarray}\label{super1}
\delta f^{\alpha(k)\dot\alpha(k)} &=& \alpha_k
\Phi^{\alpha(k)\beta\dot\alpha(k)} \zeta_\beta - \bar{\alpha}_k
\Phi^{\alpha(k)\dot\alpha(k)\dot\beta} \zeta_{\dot\beta} \nonumber
 \\
 && + \alpha'_k \Phi^{\alpha(k)\dot\alpha(k-1)} \zeta^{\dot\alpha} -
\bar{\alpha}'_k \Phi^{\alpha(k-1)\dot\alpha(k)} \zeta^\alpha,
\nonumber
 \\
\delta A &=& \alpha_0 \Phi^\alpha \zeta_\alpha - \bar{\alpha}_0
\Phi^{\dot\alpha} \zeta_{\dot\alpha} + \alpha'_0 e_{\alpha\dot\alpha}
\phi^\alpha \zeta^{\dot\alpha} - \bar{\alpha}'_0 e_{\alpha\dot\alpha}
\phi^{\dot\alpha} \zeta^\alpha,
 \\
\delta \varphi &=& \tilde{\alpha}_0 \phi^\alpha \zeta_\alpha -
\bar{\tilde{\alpha}}_0 \phi^{\dot\alpha} \zeta_{\dot\alpha}, \nonumber
\end{eqnarray}
while for the fermions respectively
\begin{eqnarray}\label{super2}
\delta \Phi^{\alpha(k+1)\dot\alpha(k)} &=& \beta_k
\Omega^{\alpha(k+1)\dot\alpha(k-1)} \zeta^{\dot\alpha} + \gamma_k
f^{\alpha(k)\dot\alpha(k)} \zeta^\alpha \nonumber
 \\
 && + \beta'_{k+1} \Omega^{\alpha(k+1)\beta\dot\alpha(k)} \zeta_\beta
+ \gamma'_{k+1} f^{\alpha(k+1)\dot\alpha(k)\dot\beta}
\zeta_{\dot\beta}, \nonumber
 \\
\delta \Phi^\alpha &=& \beta'_1 \Omega^{\alpha\beta} \zeta_\beta
 + \gamma'_1 f^{\alpha\dot\beta} \zeta_{\dot\beta} +
\beta_0 e_{\beta\dot\beta} B^{\alpha\beta} \zeta^{\dot\beta}  +
\gamma_0 A \zeta^\alpha + \hat{\gamma}_0 e^\alpha{}_{\dot\alpha}
\varphi \zeta^{\dot\alpha},
 \\
\delta \phi^\alpha &=& \tilde{\beta}_0 \pi^{\alpha\dot\alpha}
\zeta_{\dot\alpha} + \beta'_0 B^{\alpha\beta} \zeta_\beta +
\tilde{\gamma}_0 \varphi \zeta^\alpha, \nonumber
\end{eqnarray}
where $\zeta_\alpha, \zeta_{\dot{\alpha}}$ are the anticommuting
supersymmetry transformation parameters. These transformations
contain the undefined yet complex coefficients.

Using the general expressions for the variation of the bosonic
Lagrangian (\ref{varb}) we obtain:
\begin{eqnarray}
\delta {\cal L}_B &=& \sum_{k=1} (-1)^k [ 4i\alpha_k
\Phi_{\alpha(k-1)\beta\gamma\dot\alpha(k)} e^\gamma{}_{\dot\gamma}
{\cal R}^{\alpha(k-1)\dot\alpha(k-1)\dot\gamma} \zeta^\beta \nonumber
 \\
 && \qquad - 4i\alpha'_k \Phi_{\alpha(k-1)\gamma\dot\alpha(k-1)}
e^\beta{}_{\dot\beta}
{\cal R}^{\alpha(k-1)\dot\alpha(k-1)\dot\beta\dot\gamma}
\zeta_{\dot\gamma} + \dots + h.c.
\end{eqnarray}
where dots stand for the contributions of the lower spin components.
Similarly, using  the analogous expression for the fermionic
Lagrangian (\ref{varf}) we get:
\begin{eqnarray}
\delta {\cal L}_F &=& \sum_{k=1} (-1)^k [ - k \bar{\beta}_k
{\cal F}_{\alpha(k)\beta\dot\alpha(k)} e^\beta{}_{\dot\beta}
\Omega^{\alpha(k-1)\dot\alpha(k)\dot\beta} \zeta^\alpha \nonumber
 \\
 && \qquad - \bar{\gamma}_k {\cal F}_{\alpha(k)\beta\dot\alpha(k)}
e^\beta{}_{\dot\beta} (f^{\alpha(k)\dot\alpha(k)} \zeta^{\dot\beta}
+ f^{\alpha(k)\dot\alpha(k-1)\dot\beta} \zeta^{\dot\alpha}) \nonumber
 \\
 && \qquad + \bar{\beta}'_k
{\cal F}_{\alpha(k-1)\gamma\dot\alpha(k-1)} e^\gamma{}_{\dot\gamma}
\Omega^{\alpha(k-1)\dot\alpha(k-1)\dot\beta\dot\gamma}
\zeta_{\dot\beta} \nonumber
 \\
 && \qquad + \bar{\gamma}'_k
{\cal F}_{\alpha(k-1)\gamma\dot\alpha(k-1)} e^\gamma{}_{\dot\gamma}
f^{\alpha(k-1)\beta\dot\alpha(k-1)\dot\gamma} \zeta_\beta
+ \dots + h.c.
\end{eqnarray}
Now we have to find the explicit expressions for all the
coefficients $\alpha,\beta,\gamma$ such that $\delta({\cal
L}_B+{\cal L}_F)=0.$ The general technique here is essentially  the
same as the one described in \cite{BKhSZ19}, the main difference being
in the explicit form of Lagrangian parameters (\ref{parb}) and
(\ref{parf}). This produces the following results.
\begin{itemize}
\item All parameters $\alpha$ and $\gamma$ can be expressed in terms
of $\beta$:
\begin{eqnarray*}
\alpha_k &=& \frac{ik}{4}\bar{\beta}_k, \qquad
\alpha'_k = \frac{i}{4k}\bar{\beta}'_k,
 \\
\alpha_0 &=& - \frac{i}{4}\bar{\beta}_0, \quad
\tilde{\alpha}_0 \frac{i}{24}\bar{\tilde{\beta}}_0, \quad
\alpha'_0 = - \frac{i}{8}\bar{\beta}'_0,
 \\
\gamma_k &=& 2d_{k+1}\bar{\beta}_k, \qquad
\gamma'_k = 2d_k\bar{\beta}'_k,
 \\
\gamma_0 &=& - d_1\bar{\beta}_0, \quad
\tilde{\gamma}_0 = - \frac{12c_0d_1}{\tilde{a}_0}\bar{\beta}_0, \quad
\hat{\gamma}_0 = - \frac{\tilde{a}_0}{8}\beta_0.
\end{eqnarray*}
Note that these relations are purely kinematical and are the same as
in the massive case.
\item We obtain an important relation on the two dimensionful
parameters, one from bosonic sector and another one from fermionic
sector.
\begin{equation}
a_0 = c_0
\end{equation}
As it is well known, supersymmetry requires that all the members of
massive supermultiplets have the same mass. Our fields are massless
but they are characterized by the dimensionful parameters (related
with the value of the second Casimir operator of Poincare group). So
it seems natural that supersymmetry requires that these parameters
have to be related.
\item At last, we obtain a very simple (in comparison with the
massive case) solution for the remaining parameters:
\begin{eqnarray}\label{solut}
\beta_k &=& \frac{1}{\sqrt{k}}\rho, \qquad
\beta'_k = \sqrt{k}\rho', \nonumber
 \\
\beta_0 &=& \sqrt{2}\rho, \quad
\beta'_0 = 2\rho', \quad
\tilde{\beta}_0 = - \sqrt{6}\rho,
\end{eqnarray}
where
$$
\rho' = \pm \bar{\rho}.
$$
Here $\pm$ sign corresponds to that of the fermionic mass terms. Note
also, that in our multispinor formalism real (imaginary) values of
$\beta$ correspond to parity-even (parity-odd) bosonic fields. Thus we
have four independent solutions.
\end{itemize}

So we managed to find the supertransformations which leave the sum
of the bosonic and fermionic Lagrangians invariant, $\delta({\cal
L}_B+{\cal L}_F)=0.$ However, the explicit calculations of the
commutator for supertransformations (\ref{super1}), (\ref{super2})
show that their superalgebra is not closed even on-shell. To see the
reason we briefly discuss a relation between massive supermultiplets
and massless infinite spine ones. We remember that in four
dimensional Minkowski space there are two massive higher spin $N=1$
supermultiplets
$$
\left( \begin{array}{ccc} & s+\frac{1}{2} & \\
s & & s' \\ & s-\frac{1}{2} & \end{array} \right), \qquad \left(
\begin{array}{ccc} & s+1 & \\ s+\frac{1}{2} & & s+\frac{1}{2}
\\ & s' & \end{array} \right).
$$
Each of them contains four fields. The highest spin in the first
case is fermionic and the highest spin in the second case is
bosonic. In both cases the $s$ and $s'$ are integer and equal. Label
$'$ means that the  corresponding bosonic field has the opposite
parity in comparison with another bosonic field from the same
supermultiplet \cite{Zin07a}. Each multiplet is characterized
on-shell by equal number of bosonic and fermionic degrees of
freedom. As we know, the massless infinite spin fields can be
obtained from the massive one in the limit where $m \to 0$, $s \to
\infty$, $ms \to const$, therefore it is natural to consider that
the massless infinite spin supermultiplet must appear as the
analogous limit from the massive one. Moreover, the limit for the
two types of the massive supermultiplets seems to be the same since
in both cases we get infinite number of all possible helicities.
Thus to construct an infinite spin supermultiplet we need four
fields
$$
\left( \begin{array}{ccc} & \Phi_+ & \\ f_+ & & f_- \\ & \Phi_- &
\end{array} \right),
$$
where $f_+$ ($f_-$) denotes parity-even (parity-odd) boson, while
the signs of the $\Phi_\pm$ correspond to that of mass terms $d_k$
in ${\cal L}_F$ (\ref{lagf}). In particular, it means that we need
all four solutions given above (\ref{solut}) and then the complete
Lagrangian for the supermultiplet under consideration have to take
the form
\begin{eqnarray}\label{lagr}
{\cal L}= {\cal L}^{(+)}_B + {\cal L}^{(-)}_B + {\cal L}^{(+)}_F +
{\cal L}^{(-)}_F,
\end{eqnarray}
where ${\cal L}^{(\pm)}_B$ is the Lagrangian ${\cal L}_B$
(\ref{lagb}) expressed in terms of $f_{\pm}$ and corresponding
auxiliary fields $\Omega_{\pm}$, and ${\cal L}^{(\pm)}_F$
corresponds to the Lagrangian ${\cal L}_F$ (\ref{lagf}) expressed in
terms $\Phi_{\pm}$ respectively.

To simplify the presentation of the results, let us introduce the
notations for bosonic field variabels:
\begin{eqnarray}\label{Newb}
f = f_+ + if_-, &\qquad& \Omega = \Omega_+ + i\Omega_-\nonumber
\\
\bar{f} = f_+ - if_-, &\qquad& \bar{\Omega} = \Omega_+ - i\Omega_-
\end{eqnarray}
Also we introduce new fermionic variables:
\begin{eqnarray}\label{Newf}
\Phi = \Phi_+ + \Phi_-, \qquad \tilde{\Phi} = \Phi_+ - \Phi_-,
\end{eqnarray}
so that the fermionic mass terms in Lagrangian for the infinite spin
supermultiplet now have the Dirac form:
$$
\Phi_+\Phi_+ - \Phi_-\Phi_- \quad \Rightarrow \quad \tilde{\Phi}\Phi.
$$

In notations (\ref{Newb}), (\ref{Newf})  the supertransformations
can be written in a very compact form:
\begin{eqnarray}\label{ST1}
\delta f^{\alpha(k)\dot\alpha(k)} &=& \frac{i\sqrt{k}}{2}\rho
\Phi^{\alpha(k)\beta\dot\alpha(k)} \zeta_\beta +
\frac{i}{2\sqrt{k}}\rho \tilde{\Phi}^{\alpha(k-1)\dot\alpha(k)}
\zeta^\alpha, \nonumber
 \\
\delta \bar{f}^{\alpha(k)\dot\alpha(k)} &=& \frac{i\sqrt{k}}{2}\rho
\Phi^{\alpha(k)\dot\alpha(k)\dot\beta} \zeta_{\dot\beta} +
\frac{i}{2\sqrt{k}}\rho \tilde{\Phi}^{\alpha(k)\dot\alpha(k-1)}
\zeta^{\dot\alpha},
\end{eqnarray}
\begin{eqnarray}\label{ST2}
\delta \Phi^{\alpha(k+1)\dot\alpha(k)} &=& \frac{2}{\sqrt{k}}\rho
\Omega^{\alpha(k+1)\dot\alpha(k-1)} \zeta^{\dot\alpha} +
4\sqrt{k+1}d_k\rho f^{\alpha(k+1)\dot\alpha(k)\dot\beta}
\zeta_{\dot\beta}, \nonumber
 \\
\delta \tilde{\Phi}^{\alpha(k+1)\dot\alpha(k)} &=& 2\sqrt{k+1}\rho
\bar{\Omega}^{\alpha(k+1)\beta\dot\alpha(k)} \zeta_\beta +
\frac{4d_{k+1}}{\sqrt{k}}\rho \bar{f}^{\alpha(k)\dot\alpha(k)}
\zeta^\alpha
\end{eqnarray}
and similarly for the lower spin components. Here $\rho$ is the only
free (real) parameter which determines the normalization of the
superalgebra.

Direct calculations show that the algebra of these
supertransformations is indeed closed on-shell (it is instructive to
compare the structure of the results with the zero torsion conditions
(\ref{zerotor})):
\begin{eqnarray}
\frac{1}{i\rho^2} [\delta_1, \delta_2]f^{\alpha(k)\dot\alpha(k)} &=&
\Omega^{\alpha(k)\beta\dot\alpha(k-1)} \xi_\beta{}^{\dot\alpha} +
\Omega^{\alpha(k-1)\dot\alpha(k)\dot\beta} \xi^\alpha{}_{\dot\beta}
\nonumber
 \\
 && + \frac{ka_k}{2(k+2)} f^{\alpha(k)\beta\dot\alpha(k)\dot\beta}
\xi_{\beta\dot\beta} + \frac{a_{k-1}}{2k(k+1)}
f^{\alpha(k-1)\dot\alpha(k-1)} \xi^{\alpha\dot\alpha}, \nonumber
 \\
\frac{1}{i\rho^2} [\delta_1, \delta_2]f^{\alpha\dot\alpha} &=&
\Omega^{\alpha\beta} \xi_\beta{}^{\dot\alpha} +
\Omega^{\dot\alpha\dot\beta} \xi^\alpha{}_{\dot\beta}
 + \frac{a_1}{6} f^{\alpha\beta\dot\alpha\dot\beta}
\xi_{\beta\dot\beta} - \frac{a_0}{4} A \xi^{\alpha\dot\alpha},
 \\
\frac{1}{i\rho^2} [\delta_1, \delta_2] A &=& - 2e_{\beta\dot\beta}
[B^{\alpha\beta}\xi_\alpha{}^{\dot\beta} + B^{\dot\alpha\dot\beta}
\xi^\beta{}_{\dot\alpha}] - \frac{a_0}{2} f^{\alpha\dot\alpha}
\xi_{\alpha\dot\alpha}, \nonumber
 \\
\frac{1}{i\rho^2} [\delta_1, \delta_2] \varphi &=&
\pi^{\alpha\dot\alpha} \xi_{\alpha\dot\alpha}, \nonumber
\end{eqnarray}
where the translation parameter $\xi^{\alpha\dot\alpha}$ is defined by
$$
\xi^{\alpha\dot\alpha} = \zeta_1^\alpha \zeta_2^{\dot\alpha} -
\zeta_2^\alpha \zeta_1^{\dot\alpha}.
$$

Supertransformations (\ref{ST1}) and (\ref{ST2}) are the final
results connecting two bosonic and two fermionic infinite spins in
one infinite spin supermultiplet. The corresponding invariant
Lagrangian has form (\ref{lagr}) expressed in terms of new field
variables (\ref{Newb}), (\ref{Newf}).

\section{Conclusion}

We have constructed the Lagrangian formulation for the massless
infinite spin $N=1$ supermultiplet in four dimensional Minkowski
space. Such supermultiplet consists of the two bosonic (with
opposite parities) and two fermionic infinite spin fields with the
properly adjusted dimensionful parameters. We provide the gauge
invariant Lagrangian formulation for the massless infinite spin
boson and fermions which depends on one dimensionful parameter. Then
we construct the supertransformations which leave  the sum of the
four Lagrangians invariant and such that the algebra of these
transformations is closed on-shell. We note that although our
construction was based on assumption that correct massless infinite
spin supermultiplet is obtained as the special limit of higher spin
massive supermultiplet, we have derived both supertrasformations
(\ref{ST1}), (\ref{ST2}) and the invariant Lagrangian (\ref{lagr}).
The algebra of the supertransformations is closed on-shell. The
results in a whole is completely consistent with the properties of
N=1 supersymmetric theories formulated in component approach.

We want to emphasize a power and universality of the gauge invariant
approach for derivation of the Lagrangian formulation for higher
spin fields possessing the massive or dimensionful parameters. The
approach under consideration works perfectly both for massive
bosonic and fermionic field theories and for infinite spin field
theories. Also it allows to develop successfully the corresponding
supergeneralizations as it was demonstrated in the works
\cite{BKhSZ19}, \cite{BKhSZ19a} and in this work.

In this paper we constructed the supertransformation whose algebra
is closed on-shell. Such a situation is typical for component
formulation of the supersymmetric field theory where supersymmetry
is not manifest. The manifest supersymmetry is achieved in the
framework o superfield approach (see e.g. \cite{BK}). It would be
extremely interesting to develop a superfield approach to Lagrangian
formulation of the supersymmetric infinite spin field theory and
obtain off-shell supersymmetry. First step in this direction has
been made in the work \cite{BGK19} although the problem is open on a
whole.

\section*{Acknowledgments}
I.L.B and T.V.S are grateful to the RFBR grant, project No.
18-02-00153-a for partial support. Their research was also supported
in parts by Russian Ministry of Science and High Education, project
No. 3.1386.2017. Yu.M.Z is grateful to the Erwin Schr\"odinger
Institute, Vienna for the kind hospitality during the Workshop
"Higher spins and Holography", March 11 --- April 5, 2019, where this
work was completed.

\end{document}